\documentclass[aps,prl,final,letterpaper,twocolumn,showpacs]{revtex4-1}

\usepackage{graphicx,multirow,bm}
\usepackage{hyperref}
\usepackage{color}

\usepackage{amsmath,amssymb,amsfonts,amsmath}
\usepackage{epstopdf}
\usepackage{newfloat,algcompatible,algpseudocode}
\usepackage[size=small]{caption}
\usepackage{etoolbox}
\usepackage{enumerate}

\AtEndEnvironment{algorithm}{\noindent\hrulefill\par\nobreak\vskip-8pt}

\DeclareFloatingEnvironment[
    fileext=loa,
    listname=List of Algorithms,
    name=ALGORITHM,
    placement=tbhp,
]{algorithm}
\DeclareCaptionFormat{algorithms}{\vskip-15pt\hrulefill\par#1#2#3\vskip-6pt\hrulefill}
\def\arg{{\hbox{arg}}}

\def\min{{\hbox{min}}}
\def\sgn{{\hbox{sgn}}}

\def\det{{\hbox{det}}}

\def\R{\mathbf{R}}
\def\r{\mathbf{r}}

\def\BZ{\mathrm{BZ}}

\def\G{\mathbf{G}}

\def\k{\mathbf{k}}

\def\0{\mathbf{0}}

\def\dr{~\mathrm{d}\mathbf{r}}

\newcommand{\argmin}{\operatornamewithlimits{argmin}}

\newtheorem{theorem}{Theorem}

\begin{document}

\title{Compressed Wannier modes found from an $L_1$ regularized energy functional}

\author{Farzin Barekat}
\author{Ke Yin}
\author{Russel E. Caflisch}
\author{Stanley J. Osher}
\affiliation{Department of Mathematics, University of California, Los Angeles, California 90095-1555, USA} 

\author{Rongjie Lai}
\affiliation{Department of Mathematics, University of California, Irvine, California 92697-3875, USA} 

\author{Vidvuds Ozoli\c{n}\v{s}}
\affiliation{Department of Materials Science and Engineering, University of California, Los Angeles, California 90095-1595, USA} 

\begin{abstract}
We propose a method for calculating Wannier functions of periodic solids directly from a modified variational principle for the energy, subject to the requirement that the Wannier functions are orthogonal to all their translations ("shift-orthogonality"). Localization is achieved by adding an $L_1$ regularization term to the energy functional. This approach results in "compressed" Wannier modes with compact support, where one parameter $\mu$ controls the trade-off between the accuracy of the total energy and the size of the support of the Wannier modes. Efficient algorithms for shift-orthogonalization and solution of the variational minimization problem are demonstrated. 
\end{abstract}

\date{\today} 

\maketitle

Electronic states in periodic crystals are usually discussed in terms of Bloch waves of definite crystal momentum and energy. An alternative description in terms of spatially localized functions was introduced by Gregory Wannier \cite{Wannier1937} and further developed in \cite{Kohn1959,Wannier1960,Blount1962,DesCloizeaux1963}. These so-called Wannier functions are associated with lattice sites, are translational images of each other, and can be chosen to be real and exponentially localized in conventional (i.e., topologically trivial) insulators \cite{Brouder2007}. Even though the Wannier functions are not the eigenstates of the crystal Hamiltonian, they represent a convenient description of the electronic states for understanding such phenomena as electric polarization \cite{Resta1994}, orbital magnetization \cite{Thonhauser2005}, nontrivial insulating states \cite{Resta2011} and range of electronic interactions in condensed matter \cite{Prodan2005}. Wannier functions can also be used to increase speed and accuracy of computations. For instance, they can be used to interpolate the electronic wave functions and band structure throughout the Brillouin zone whenever a very large number of $\k$ points is needed, such as when calculating electron-phonon scattering rates \cite{Giustino2007}.

Wannier functions are unitary transformations of Bloch waves with different crystal momenta and are usually obtained by optimizing a suitably chosen localization functional. A particularly successful choice was introduced by Marzari and Vanderbilt \cite{Marzari1997}, in which one minimizes the spread (second moment) of the Wannier functions, resulting in maximally localized Wannier functions (MLWF). Due to the non-convexity of the target functional and constraints, a reasonable initial guess is usually needed to avoid local minima corresponding to poorly localized, complex Wannier functions \cite{Marzari2012}.

It is well understood that for insulators Wannier functions satisfy the minimum principle for the total energy subject to the constraint of orthogonality to all their translations by lattice vectors; we refer to this as shift-orthogonality. The corresponding variational principle was formulated by Koster \cite{Koster1953} and used by Kohn \cite{Kohn1973} in his variational Wannier function approach, but it has been seldom used in practice with general bases \cite{Pederson1987}. 

In this paper, we show that localized Wannier modes can be obtained directly from an $L_1$ regularized variational principle without ever calculating crystal eigenstates in the Bloch representation. These ideas generalize earlier work \cite{Ozolins2013,Ozolins2014} to systems with translational symmetry. Our approach is well-defined for both insulating and metallic systems, with one parameter providing a systematically controllable trade-off between the accuracy of the total energy and the localization degree of the regularized ("compressed") Wannier modes. We also introduce efficient numerical methods to solve the associated constrained variational problem.

For simplicity, we assume that the problem permits real-valued Wannier functions. Following general practice, we label the Wannier functions $\psi^n_{\R} (\r) \equiv \psi^n (\r - \R)$ by a band index $n$ and lattice site $\R$. The  $L^1$ regularized energy functional introduced in \cite{Ozolins2013} is written as 
\begin{equation}\label{eq:modified_functional}
\mathcal{J}(\psi):=\frac{1}{\mu} \|\psi\|_1+\langle \psi | \hat{H} | \psi \rangle,
\end{equation}
where the $L_1$ norm of a function is defined as $\|\psi\|_1=\int |\psi(\r)|\,\mathrm{d}\r$. The effect of the $L_1$ term is to localize the solutions, and the parameter $\mu$ controls the trade-off between sparsity and accuracy: larger values of $\mu$ give solutions that better minimize the total energy at the expense of more extended Wannier functions, while a smaller $\mu$ gives highly localized wave functions at the expense of larger errors in the calculated energies. Furthermore, due to the properties of the $L_1$ term, the functions that minimize Eq.~\eqref{eq:modified_functional} have compact support, i.e. they are nonzero only in a finite spatial region. Since the functional \eqref{eq:modified_functional} is convex, efficient numerical minimization methods can be devised.

Our proposed scheme defines compactly supported Wannier modes recursively by minimizing  $\mathcal{J}(\psi)$  subject to shift-orthogonality and normalization constraints: 
\begin{equation}
\label{eq:Wannier}
\begin{cases}
\psi^1=\underset{\psi}{\arg \min}\;\mathcal{J}(\psi) & \hbox{s.t.} \;\; \langle\psi_{\R}| \psi_{\R' } \rangle=\delta_{\R\R'} \\
\psi^k=\underset{\psi}{\arg \min}\; \mathcal{J}(\psi) & \hbox{s.t.} \;\; \langle\psi_{\R}| \psi_{\R'} \rangle=\delta_{\R\R'} \\ 
								&\textrm{and} \;\; \langle \psi_\R | \psi^i_{\R'} \rangle=0 \; \textrm{for}\; i<k.
\end{cases}
\end{equation}
This generalizes to nonzero crystal potentials the approach used in \cite{Ozolins2014} to construct the compressed plane wave (CPW) bases for the Laplacian. A key advantage of our scheme is that one parameter $\mu$ controls both the physical accuracy and the spatial extent, while not requiring any physical intuition about the properties of the solution. In other words, the Wannier functions are nonzero only in those regions that are required to achieve a given accuracy for the total energy and are zero everywhere else.  The Cauchy--Schwarz inequality guarantees that the difference between \eqref{eq:modified_functional} and the true energy functional is bounded from above by a constant multiple of $\|\psi\|_2$. Hence, the solutions to the variational problem involving \eqref{eq:modified_functional} provide an accurate, systematically controllable approximation to the true total energy of the system \cite{Ozolins2013}. Fully self-consistent calculations can be performed using Wannier functions, without any reference to the Bloch waves and Brillouin zones. 

In what follows, we describe efficient algorithms for solving \eqref{eq:Wannier}.  We choose a supercell $\Omega$ defined by three lattice vectors 
\begin{equation}
{\R}^\text{SC}_\alpha = \sum_{\beta=1}^3 L_{\alpha\beta} {\R}_\beta, 
\end{equation}
where $L_{\alpha\beta}$ is a $3 \times 3$ nonsingular matrix with integer elements, and ${\R}_\beta$ are the unit cell vectors of the primitive lattice. The Hamiltonian $\hat{H} = - \frac{1}{2} \Delta + V(\r)$ has the periodicity of the primitive lattice:
\begin{eqnarray}
\label{eq:Hperiodic}
V(\mathbf{r}) = V ( \r + \R ), \\
\R = \sum_{\alpha=1}^3 n_\alpha \mathbf{R}_\alpha, \quad n_\alpha \in \mathbb{Z}.
\end{eqnarray}
For the Wannier modes in \eqref{eq:Wannier}, we impose periodic boundary conditions with respect to the supercell:
\begin{eqnarray}
\label{eq:PBC}
\psi ({\r}) = \psi ( \r +  \R_\text{S} ) \\
{\R}_\text{S} = \sum_\alpha n_\alpha \mathbf{R}^\text{SC}_\alpha, \quad n_\alpha \in \mathbb{Z}.
\end{eqnarray}
For physical accuracy, the supercell $\R^\text{SC}_\alpha$ should be chosen big enough to allow the compressed Wannier modes to decay to zero within the range of the supercell, although this is not necessary for the numerical algorithm to work. 

We also introduce the primitive reciprocal lattice ${\bf Q}_\alpha$ such that ${\bf Q}_\alpha {\bf R}_\beta = 2 \pi \delta_{\alpha\beta}$.
The Fourier expansion of a supercell periodic Wannier mode $\psi$ will contain only plane waves with wave vectors $\k+\G$,
where $\k$ belongs to the first Brillouin zone of the primitive lattice and $\G = \sum_\alpha m_\alpha \mathbf{Q}_\alpha \;\; (m_\alpha \in \mathbb{Z})$ is a reciprocal lattice vector. Fourier expansion of Wannier mode $\psi$ includes all plane waves below a certain kinetic energy cutoff $E_\mathrm{max}$:
\begin{equation}
\label{eq:Emax}
\frac{1}{2} |\k + \G |^2 \leq E_\mathrm{max},
\end{equation}
and can be written as
\begin{equation}
\label{eq:Fourier}
\psi( \mathbf{r} ) = \sum_\k \sum_\G \tilde{\psi} (\k+\G) e^{i(\k+\G)\r} \equiv \sum_\k u_{\k} (\r) e^{i \k \r},
\end{equation}
where we have defined a cell-periodic Bloch function
\begin{equation}
\label{eq:Bloch}
u_{\k} (\r) = \sum_\G \tilde{\psi} (\k+\G)  e^{i \G \r}.
\end{equation}
The inverse Fourier transform is given by
\begin{equation}
\tilde{\psi}(\k+\G)=\frac{1}{| \Omega |}\int_\Omega \psi(\r)e^{-i(\k+\G)\r}\dr,
\label{equation:inverseFourier}
\end{equation}
where the integral extends over the supercell and $|\Omega|$ is the supercell volume.

{\it Shift-orthogonality:\/} One of the key steps for \eqref{eq:Wannier} is ensuring that the solution $\psi^n$ is orthogonal to its own translations by all primitive lattice vectors $\R$, as well as orthogonal to all translations of the lower Wannier modes $\psi^1 \ldots \psi^{n-1}$. We say that function $\psi(\r)$ is shift-orthogonal if and only if 
\begin{equation}
\langle \psi_{\R'} | \psi_\R \rangle \equiv \int \psi(\r - \R') \psi(\r - \R) \dr = \delta_{\R'\R}
\label{equation:shiftOrthogonalDefinition}
\end{equation}
holds for all lattice vectors $\R$ and $\R'$. The nonlinear Lagrangian method used to enforce \eqref{equation:shiftOrthogonalDefinition} in \cite{Ozolins2014} is too slow in this context, and here we detail a faster approach adapted from computational harmonic analysis \cite{Mallat1999}. 

Given a supercell periodic function $f$, the objective is to find the projection of $f$ to the set of shift orthogonal functions (see also \cite{BarekatThesis}). In other words, we need to solve the following minimization problem:
\begin{equation}
\hat{\Pi} f := \argmin_{\psi} \| f - \psi \|_2 \quad \hbox{s.t.} \quad \langle \psi_{\R'} | \psi_{\R} \rangle = \delta_{\R'\R}.
\label{equation:SOproblem}
\end{equation}
Note that $\hat{\Pi} f$ is not necessarily unique and the set of shift-orthogonal functions is not a vector space because a sum of two shift-orthogonal functions may not be shift-orthogonal. However, if two shift-orthogonal functions $f$ and $g$ are orthogonal to all shifts of each other, any normalized linear combination of them will also be shift-orthogonal. This property allows to design efficient iterative update algorithms for \eqref{eq:Wannier}.


The following theorem is well known in the wavelet community (e.g., see Eq. 7.19 in \cite{Mallat1999}). For completeness, we provide the proof of the theorem in the appendix.
\begin{theorem}
Supercell-periodic function $\psi(\r)$ is shift-orthogonal if and only if
\[ \sum_{\G}|\tilde{\psi}(\k+\G)|^2=\frac{1}{N|\Omega|} \quad \forall~ \k\in \BZ, \]
where $N=\det(L)$ is the number of primitive cells inside the supercell.
\label{theorem:equivalence}
\end{theorem}

A derivation similar to what is used for Theorem \eqref{theorem:equivalence} yields the following theorem:
\begin{theorem}
For two supercell-periodic functions $\psi(\r)$ and $\phi(\r)$,
\[ \langle \psi_\R | \phi_{\R'} \rangle = 0 \quad \forall~\R,\R' \in \Omega \]
if and only if
\[ \sum_{\G}\tilde{\psi}^*(\k+\G) \tilde{\phi}(\k+\G)=0 \quad \forall~ \k \in \BZ. \]
\label{theorem:equivalence2}
\end{theorem}

It is seen that these shift-orthogonality conditions amount to orthonormalization imposed on the Bloch functions $u_{\k} (\r)$ \eqref{eq:Bloch}.

%

Theorem \ref{theorem:equivalence} and Parseval's identity $\|f-\psi\|_2=|\Omega| \|\tilde{f}-\tilde{\psi}\|_2$ yield a straigtforward algorithm for obtaining the solution to problem \eqref{equation:SOproblem}. This algorithm has several important features. First, it has computational complexity of $M\log(M)$ where $M$ is the number of Fourier coefficients used to represent function $f$. Second, it is  parallelizable over both $\k$ and $\G$. Finally, for a real valued input function $f$, the algorithm outputs a real valued projection $\hat{\Pi} f$. As mentioned earlier, the solution to \eqref{equation:SOproblem} is not unique if $\sum_{\G}|\tilde{\psi}(\k+\G)|^2=0$ for some $\k$. In these situations, we choose the solution corresponding to the lowest frequency, i.e. $\tilde{\psi}(\k)=\frac{1}{N |\Omega|}$.

Next suppose that supercell periodic functions $f$ and $g^1,\ldots,g^n$ are given. Theorem \ref{theorem:equivalence2} yields an algorithm similar to the one discussed above that finds a shift-orthogonal projection that is also perpendicular to all translations of  $g^1,\ldots,g^n$:
\begin{align}\label{equation:SOproblemV2}
&\hat{\Pi}_{\{g^1,\ldots,g^n\}^\perp} f := \argmin_{\psi} \|f-\psi\|_2  \quad \hbox{s.t. } \\
& \hspace{1cm}\begin{cases}
\hbox{$\psi$ is shift-orthogonal, and}  \\
\langle \psi_{\R} | g^i_{\R'} \rangle=0 \quad \hbox{for } \forall \, \R,\R' \in \Omega \;\; \text{and} \;\; i=1,\ldots,n.
\end{cases} \notag
\end{align}


{\it Computing Wannier modes:\/}
Wannier functions as in \eqref{eq:Wannier} are minimizers of $\mathcal{J}(\psi)$ subject to shift-orthogonality constraints. 
However, minimization of $\mathcal{J}(\psi)$ cannot be done efficiently using conventional quadratic optimization techniques due to the discontinuous behavior of the derivative of the $L_1$ term at $\psi = 0$ and the non-convex constraints. Efficient numerical methods for such problems are based on the Bregman iteration \cite{Osher2005,Yin2008}. Here, we use the split Bregman approach of \cite{Goldstein2009}, which treats the $L^1$ term by introducing an additional variable $v$ with a quadratically constraint to approach the solution, as shown in Algorithm \ref{algorithm:bregman_L1}. The main advantage of this approach is that the minimization of the quadratic functional $\langle \psi | \hat{H} | \psi \rangle$ is separated from the minimization of the $L^1$ term, allowing use of highly efficient quadratic optimization algorithms for the former.

\begin{algorithm}[h]
\caption{Algorithm for finding a level-$k$ compressed Wannier mode by split Begman iteration}
\label{algorithm:bregman_L1}
\begin{algorithmic}[1]
\Statex \textbf{Input: }$\hat{{H}}$, $\left\{ \psi^{j}:j=1,\ldots,k-1\right\} $
(empty if $k=1$), $\lambda, \gamma, \mu$ 

\Statex \textbf{Output: } {$\psi^k(\r)$}

\State {\bf{Initialize}}: both $u(\r),v(\r)$ are norm-1 random functions defined on $\Omega$. $b(\r)=c(\r)=0$;
\While {``not converged'' }
	\State $\displaystyle \psi=\underset{\psi}{\argmin}\langle\psi|\hat{{H}}|\psi\rangle+\frac{\lambda}{2}\|\psi-u+b\|_{2}^{2}+\frac{{\gamma}}{2}\|\psi-v+c\|_{2}^{2}$\;
	\State $\displaystyle u = \hat{\Pi}_{\{\psi^1,\ldots,\psi^{k-1}\}^\perp} (\psi+b) $;
	\State $\displaystyle v=\underset{v}{\text{{argmin}}}\frac{{1}}{\mu}\|v\|_{1}+\frac{{\gamma}}{2}\|\psi-v+c\|_{2}^{2}$;
	\State $b=\psi-u+b$;
	\State $c=\psi-v+c$;
\EndWhile
\State {\bf{return}} $\psi^k(\r)= \psi$. 
\end{algorithmic}
\end{algorithm}


In Algorithm \ref{algorithm:bregman_L1}, $\lambda,\gamma$ are chosen such that $\hat{H}+\lambda+\gamma$ is positive definite, and $\mu$ is chosen such that $\gamma \mu\gg \frac{1}{\sqrt{| \Omega |}}$. Line 3 is equivalent to solving the elliptic equation \eqref{eq:non_linear_elliptic}
\begin{equation}
\label{eq:non_linear_elliptic}
(\hat{H}+\lambda+\gamma) \psi(\r) = \lambda [ u(\r)-b(\r)] + \gamma [v(\r)-c(\r)],
\end{equation}
which can be solved by the preconditioned conjugate gradient method, with the preconditioner given by the inverse of a linear elliptic operator
\[
\left( -\frac{1}{2}\nabla^2 + \lambda + \gamma \right )^{-1}.
\]
In this work, we implement the inverse by a fast Poisson solver. In practice, problem \eqref{eq:non_linear_elliptic} does not need to be solved exactly and a few iterations per cycle are sufficient. Line 4 is solved by the algorithms described for \eqref{equation:SOproblem} when $k=1$ and by \eqref{equation:SOproblemV2} when $k>1$. Line 5 is solved by a component-wise soft thresholding operation: 
\[ v(\r)=\sgn \left( \psi(\r)+c(\r) \right) \max \left( 0,|\psi(\r)+c(\r)|-\frac{1}{\gamma\mu} \right). \]

\begin{figure}[htbp]
\includegraphics[width=1\linewidth]{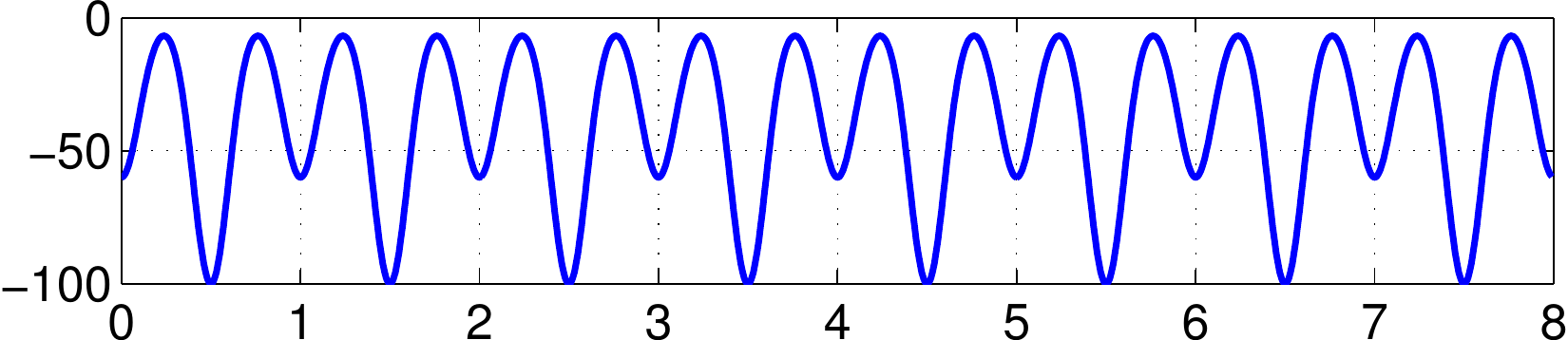}
\includegraphics[width=1\linewidth]{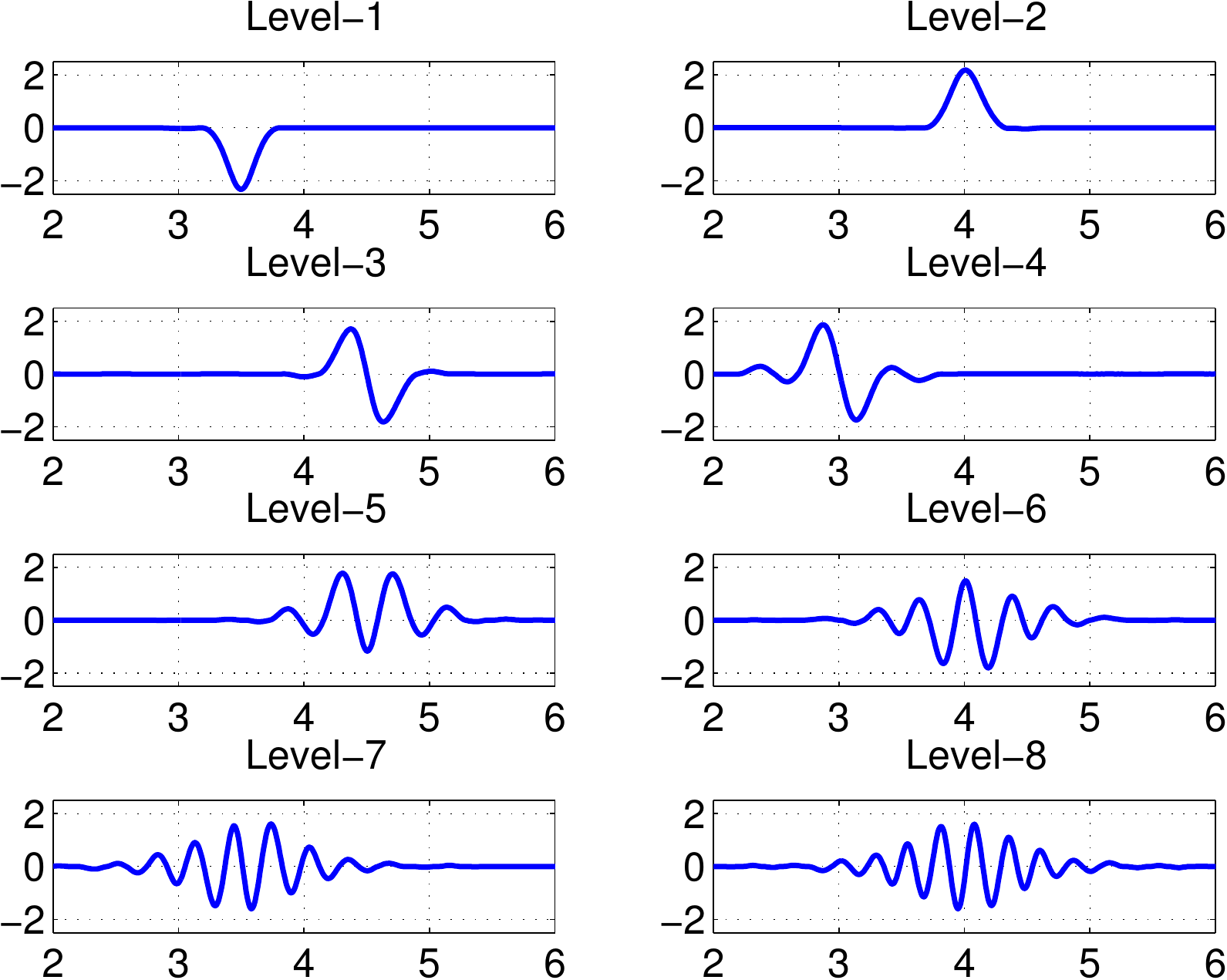}

\caption[]{From top to bottom: Model potential function (top) and its compressed Wannier modes at levels 1-8 (bottom). }
\label{fig:Wannier-Functions}
\end{figure}

As an illustration, we find compressed Wannier modes for a one-dimensional system with lattice parameter $a$ and Hamiltonian $\hat{{H}}=-\frac{{1}}{2}\nabla^2+V(x)$, where $V$ is a superposition of inverted Gaussians of two different depths:
\[
V(x)=-\sum_{j=-\infty}^{\infty} \sum_{m=1}^2 V_{m} \exp \left[ -\frac{(x-x_m-ja)^{2}}{2\sigma^{2}} \right].
\]
We choose $a=1$, $x_1=0$, $V_1=60$ and $x_2=a/2$, $V_2=1 \times 100$; the resulting potential $V(x)$ is shown in Figure \ref{fig:Wannier-Functions}. The lowest 8 Wannier modes are constructed following Algorithm \ref{algorithm:bregman_L1} using a supercell of length $L=8$ and parameters $\lambda=\gamma=10^3, \mu=10/\sqrt{L}$. The results are shown in Figure \ref{fig:Wannier-Functions}. Observe that Wannier modes of levels 1 and 3 are located within the the deep wells and level 2 is located in the shallow well, corresponding to "semi-core states" with flat bands. Higher levels spread over the two types of wells, which suggests that they belong to the continuous spectrum. Adaptively changing $\mu$ in front of the $L^1$ term inversely proportional to the total energy can limit their support.  We note that the Wannier modes in all cases are either  symmetric or antisymmetric, i.e. they constitute irreducible representations of the symmetry group of the underlying potential. It remains to be seen whether similar property is preserved in higher dimensions.

\begin{figure}[htbp]
\includegraphics[width=1\linewidth]{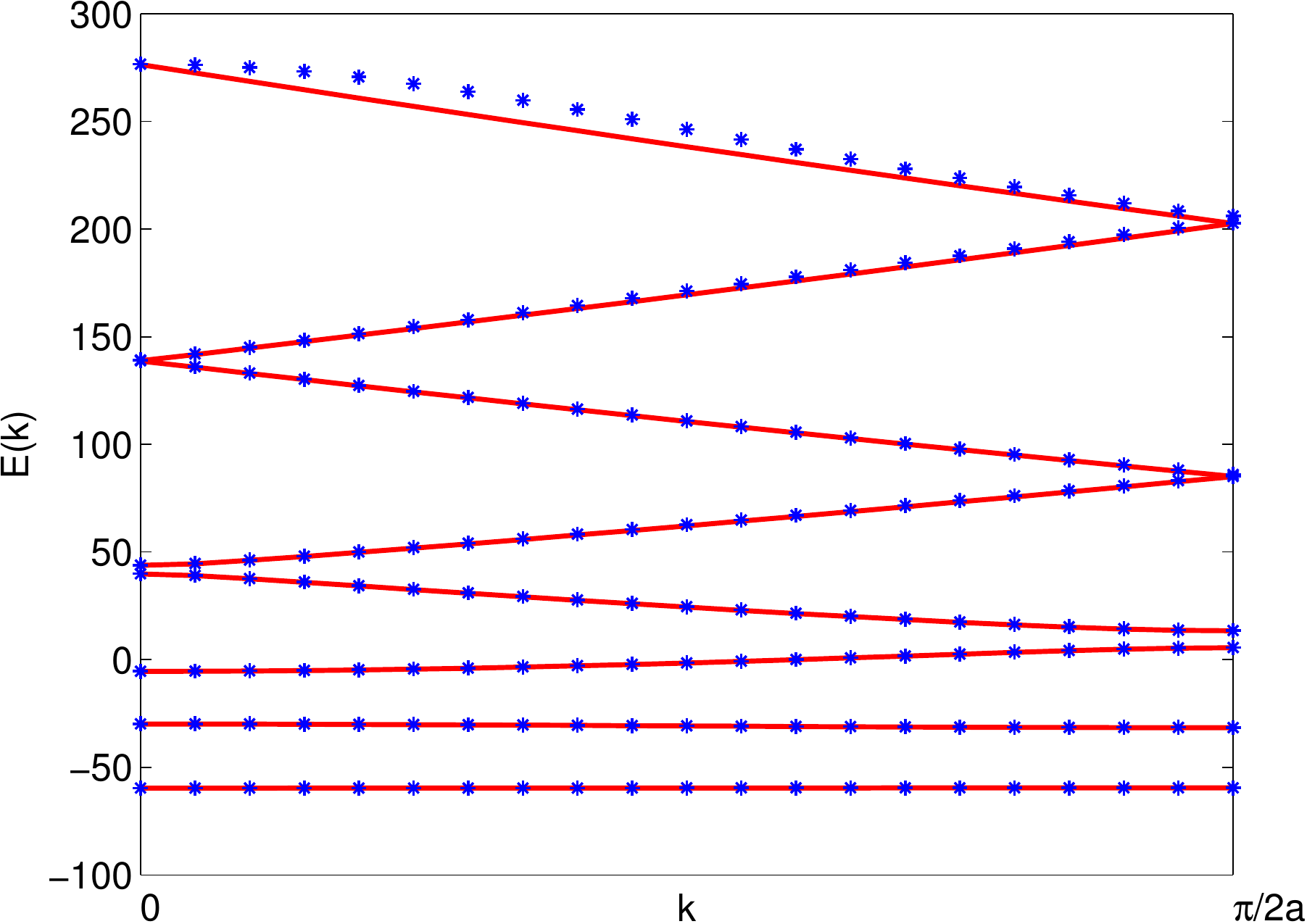}
\caption[]{Eigenvalue dispersion for bands 1-8 calculated by exact diagonalization (continuous line) and by using the lowest 8 Wannier modes (filled circles).}
\label{fig:spaghetti}
\end{figure}

The calculated eigenvalue dispersion for bands 1-8 is shown in Figure~\ref{fig:spaghetti} for exact diagonalization (continuous line) and for subspace diagonalization using the lowest 8 Wannier modes (filled circles). The former are calculated as the eigenvalues of the subspace Hamiltonian,
\begin{equation}
h^{nm}(\k) = \sum_{\R} e^{-i\k\R} \langle \psi^n_\mathbf{0} | \hat{H} | \psi^n_{\R} \rangle.
\end{equation}
We see that the agreement is perfect, except for small deviation in the highest band, which is due to the limited number of Wannier modes in use.

In conclusion, we have introduced an approach to obtaining compactly supported Wannier modes directly from an $L_1$ regularized variational principle for the total energy. Our approach does not require calculation of the Bloch states with a subsequent minimization of a nonconvex localization functional and therefore is expected to be more robust. The proposed numerical algorithms are logically straightforward and simple to implement in existing density-functional theory (DFT) codes with Brillouin zone sampling. Indeed, the key step of Algorithm~\ref{algorithm:bregman_L1} involves iterative solution of Eq.~\eqref{eq:non_linear_elliptic}, which in turn requires evaluations of $\hat{H} \psi$. Using the decomposition of $\psi$ into Bloch functions according to \eqref{eq:Bloch}, we can write
\begin{equation}
\label{eq:Hxpsi}
\hat{H} \psi = \sum_{\k} e^{i\k\r} \hat{H}_{\k} u_{\k} (\r),
\end{equation}
where $\hat{H}_{\k} \equiv e^{-i\k\r} \hat{H} e^{i\k\r}$. Routines for calculating $\hat{H}_{\k} u_{\k} (\r)$ are already implemented in codes based on the Bloch theorem, and \eqref{eq:Hxpsi} can be evaluated by a simple summation or Fourier transform over $\k$ in the Brillouin zone. Hence, the computational complexity of the proposed approach is similar to that of conventional Bloch function methods and can be used directly in self-consistent DFT calculations. We also hypothesize that $L_1$ regularized Wannier modes will be useful for beyond-DFT approaches that can benefit from the finite range of electronic states, such as screened exchange and quantum Monte Carlo methods.

V.O. was supported by the  National Science Foundation under Award No. DMR-1106024 and used computing resources at the National Energy Research Scientific  Computing Center, which is supported by the US DOE under  Contract No. DE-AC02-05CH11231. The research of R.C. is partially supported by the US DOE under  Contract No. DE-FG02-05ER25710. The research of S.O. was supported by the Office of Naval Research (Grant N00014-11-1-719). We acknowledge Dr. J. C. Budich, who pointed out the equivalence of shift-orthogonality conditions for the Wannier functions and the orthonormality of their Bloch functions.

\appendix
\section{Proof of shift orthogonality theorems}
\label{section:SOcriteria}

Here we present a proof for Theorem \ref{theorem:equivalence}. For a given function $h$ on $\Omega$, define its sampling function at lattice points of $\Omega$ by
\begin{equation} h_d(\r)=\sum_{\R\in \Omega}h(\r)\delta_{\R}(\r),  \label{equation:definitionHd}\end{equation}
where $\delta_{\R}(\r)$ is the Dirac delta function $\delta(\r-\R)$. Let $N$ denote the number of primitive cells inside the supercell (i.e. $N=\det(L)$). Note that 
\begin{equation}
\tilde{h}_d(\k+\G)= N\sum_{\G'} \tilde{h}(\k+\G').
\label{equation:hHat}\end{equation}
To see this, observe that
\begin{align*}
~~&\tilde{h}_d(\k+\G) \\
=&  \sum_{\R\in \Omega}|\Omega| (\tilde{h}*\tilde{\delta}_{\R})(\k+\G) \\
=&  \sum_{\R\in \Omega}\sum_{\G'} \sum_{\k'\in \BZ}|\Omega| \tilde{h}(\k'+\G')\frac{e^{-i (\k+\G-(\k'+\G'))\R}}{|\Omega|} \\
=&  \sum_{\G'} \sum_{\k'\in \BZ} \tilde{h}(\k'+\G') \sum_{\R\in \Omega} e^{-i (\k-\k')\R}   \\
=&  \sum_{\G'} \sum_{\k'\in \BZ} \tilde{h}(\k'+\G') N {\mathbf{1}}_{\{\k'=\k\}}  \\
=&  N\sum_{\G'} \tilde{h}(\k+\G'). 
\end{align*} 

In view of \eqref{equation:shiftOrthogonalDefinition}, supercell-periodic function $\psi(\r)$ is shift-orthogonal if and only if for all $\R'\in \Omega$:
\begin{align}
\delta_{\R'\0} &= \langle \psi(\r-\R'), \psi(\r) \rangle = \int_\Omega \psi^*(\r-\R')\psi(x)= \notag  \\
&=\int_\Omega \mathring{\psi}(\R'-\r) \psi(\r) =(\mathring{\psi}*\psi)(\R'),  \label{equation:delR0}
\end{align}
where $\mathring{\psi}$ is defined by $\mathring{\psi}(\r)=\psi^*(-\r)$.

Now, let $h(\r)=(\mathring{\psi}*\psi)(\r)$. In view of \eqref{equation:definitionHd} and \eqref{equation:delR0}, $\psi(\r)$ is shift-orthogonal if and only if 
\[ h_d(\r)=\delta(\r). \]
Taking Fourier transform from both sides, using \eqref{equation:hHat} and using the fact that Fourier transform of $(\mathring{\psi}*\psi)(\r)$ is $|\tilde{\psi}(\k+\G)|^2$ yields that $\psi(\r)$ is shift-orthogonal if and only if  for all $\k\in \BZ$,
\[ N \sum_{\G'} |\tilde{\psi}(\k+\G')|^2=\frac{1}{|\Omega|}. \]

%

\end{document}